\documentclass[conference]{IEEEtran}

\pdfoutput=1

\usepackage[T1]{fontenc}
\usepackage[utf8]{inputenc}
\usepackage{lmodern}

\usepackage{amsmath,amssymb,amsfonts}

\usepackage{cite}
\usepackage[english]{babel}
\usepackage{color}
\usepackage{booktabs}
\usepackage[normalem]{ulem}
\usepackage{xspace}
\usepackage{placeins}
\usepackage{graphicx}
\usepackage{inconsolata} 

\usepackage{caption}
\usepackage{listings}
\definecolor{mygray}{rgb}{0.5,0.5,0.5}
\definecolor{myblue}{rgb}{0,0,1}
\definecolor{mygreen}{rgb}{0,0.5,0}
\definecolor{myred}{rgb}{0.64,0.08,0.08}
\let\othelstnumber=\thelstnumber
\def\createlinenumber#1#2{
    \edef\thelstnumber{%
        \unexpanded{%
            \ifnum#1=\value{lstnumber}\relax
              #2%
            \else}%
        \expandafter\unexpanded\expandafter{\thelstnumber\othelstnumber\fi}%
    }
    \ifx\othelstnumber=\relax\else
      \let\othelstnumber\relax
    \fi
}

\usepackage[hidelinks,bookmarks=false]{hyperref}

\newcommand{\pdftitle}{Autofolding for Source Code Summarization}

\hypersetup{
  pdftitle={\pdftitle},
  pdfauthor={},
  plainpages=false,
}
\addto\extrasenglish{}
\addto\extrasenglish{}
\addto\extrasenglish{}

\renewcommand{\baselinestretch}{1.0}
\newcommand{\cut}[1]{}

\newcommand{\cf}{\hbox{{cf.}}\xspace}

\newcommand{\etal}{\hbox{{et al.}}\xspace}
\newcommand{\eg}{\hbox{{e.g.}}\xspace}
\newcommand{\ie}{\hbox{{i.e.}}\xspace}

\newcommand{\etc}{\hbox{{etc.}}\xspace}

\newcommand{\deq}{\mathrel{\mathop:}=}

\newcommand{\wB}{\mathbf{w}}
\newcommand{\zB}{\mathbf{z}}
\newcommand{\bu}{\mathbf{u}}
\newcommand{\bv}{\mathbf{v}}

\newlength{\emstr}
\newcommand{\boldpara}[1]{%
\smallskip%
\par\noindent\textbf{\textit{#1}}\hspace{\emstr}
}%

\usepackage{enumitem}
\setitemize{noitemsep,topsep=0pt,parsep=0pt,partopsep=0pt,leftmargin=*}
\setenumerate{noitemsep,topsep=0pt,parsep=0pt,partopsep=0pt,leftmargin=*}
\setdescription{noitemsep,topsep=0pt,parsep=0pt,partopsep=0pt,leftmargin=*}

\begin{document}

\title{\pdftitle}

\author{\IEEEauthorblockN{
Jaroslav Fowkes\IEEEauthorrefmark{1},
Pankajan Chanthirasegaran\IEEEauthorrefmark{1},
 Razvan Ranca\IEEEauthorrefmark{2}, \\
 Miltiadis Allamanis\IEEEauthorrefmark{1},
Mirella Lapata\IEEEauthorrefmark{1} and
 Charles Sutton\IEEEauthorrefmark{1}}
\IEEEauthorblockA{\IEEEauthorrefmark{1}School of Informatics, University of Edinburgh, Edinburgh, EH8 9AB, UK \\
\{jaroslav.fowkes, pchanthi, m.allamanis, csutton\}@ed.ac.uk; mlap@inf.ed.ac.uk}
\IEEEauthorblockA{\IEEEauthorrefmark{2}Tractable, Oval Office, 11-12 The Oval, London, E2 9DT, UK \\
razvan@tractable.io}
}

\maketitle

\begin{abstract}
Developers spend much of their time reading and browsing source code, raising new opportunities for summarization methods. Indeed, modern code editors provide code \emph{folding}, which allows one to selectively hide blocks of code. However this is impractical to use as folding decisions must be made manually or based on simple rules. We introduce the \emph{autofolding problem}, which is to automatically create a code summary by folding less informative code regions. We present a novel solution by formulating the problem as a sequence of AST folding decisions, leveraging a scoped topic model for code tokens. On an annotated set of popular open source projects, we show that our summarizer outperforms simpler baselines, yielding a 28\% error reduction. Furthermore, we find through a case study that our summarizer is strongly preferred by experienced developers. More broadly, we hope this work will aid program comprehension by turning code folding into a usable and valuable tool.

\end{abstract}


\section{Introduction}\label{intro}
Engineering large software systems presents many challenges due to the
inherent complexity of software. 
Because of this complexity,
programmers tend to spend more time reading and browsing code 
than actually writing it \cite{latoza2006maintaining,ko2006exploratory}.
Despite much research
\cite{storey2005theories}, there is still a large need for better tools 
that aid program comprehension, thereby reducing the cost of software development.

A key insight is that source code is written to be understood not only by machines,
but also by humans. Programmers devote significant time and attention to writing
their code in an idiomatic and intuitive way that can be easily
understood by others --- \emph{source code is a means of human communication.}
This fact raises the intriguing possibility that technology
from the natural language processing (NLP) community can be 
adapted to help developers make sense of large
repositories of code. 
Often during development and maintenance, developers skim the code in order to quickly understand it \cite{starke2009searching}.
A good \emph{summary} of the source code aims to support this use case:
by eliding less-important details, a summary can be easier to read quickly 
and help the developer to gain a high-level conceptual understanding of the code.

Source code summarization has potential for valuable applications
in many software engineering tasks, such as: (a) \emph{Understanding new code bases.}
Often developers need to quickly familiarize themselves with 
the core parts of a large code base. This can happen when a developer is joining
an existing project,
or when a developer is evaluating whether to use a new software library.
(b) \emph{Code reviews.} Reviewers need to quickly understand the key changes 
before reviewing the details. (c) \emph{Locating relevant code segments.} 
During program maintenance, developers often skim  code, reading only a couple lines at a time, while 
searching for a code region of interest \cite{starke2009searching}.

For this reason, many code editors include a feature called \emph{code folding},
which allows developers to selectively display or hide blocks of source code. This
feature is commonly supported in editors and is familiar to developers \cite{hendrix1998visual, kullbach2001folding, rugaber2008conceptual}.
But in current Integrated Development Environments (IDEs), folding quickly becomes
impractical because the folding decisions must be done manually by the programmer,
or based on simple rules, such as folding code blocks based on depth \cite{bragdon2010code}, that some IDEs take automatically.
This creates an obvious chicken-and-egg problem, because the developer must 
already understand the source file to decide what should be folded. 

In this paper, we propose that code folding can be a valuable tool
for aiding program comprehension, provided that folding decisions
are made automatically based on the \emph{code's content}.
We consider the \emph{autofolding problem}, in which the goal is to
automatically create a code summary by folding non-essential code elements
that are not useful on first viewing.
To our knowledge, we are the first to systematically study
and quantitatively compare different methods for the autofolding problem.
An illustrative example is shown in \autoref{fig:ex}. To any Java developer the function of the
\lstinline+StatusLine+ constructor and the \lstinline+clone+, \lstinline+getCode+, \lstinline+getReason+
and \lstinline+toString+ methods are obvious even without seeing their method bodies.
One possible summary of this source file is
shown in \autoref{fig:ex2}. 

The key  problem in  content-based autofolding 
is to determine which tokens in a file are most representative
of its content.  We compare two different content models for this task: 
a simple vector space model (VSM) and a topic model that, 
building on work in NLP summarization \cite{haghighi2009exploring},
endows different scopes (files, projects, and the corpus)
with separate topics, allowing the model to separate out those tokens 
that are used most often in a particular file.
We find that the summaries from the topic model are significantly better 
than those from the VSM.

Previous work in code summarization has considered summarization using: 
(a) program slicing (\ie hiding irrelevant lines of code for a chosen program path) \cite{silva2012vocabulary,khoo2008path}; 
(b) natural language paraphrases \cite{sridhara10summary,moreno2013automatic}; 
(c) short lists of keywords \cite{haiduc2010supporting,haiduc2010automated,eddy2013evaluating,mcburney2014improving};
or (d) (potentially discontiguous) lines of code that match a user's query \cite{ying2013code}. 
In contrast, our work is based on the
idea that an effective summary can be obtained by carefully folding the original file 
--- summarizing code \emph{with code}. 
Our main contributions in this paper are: 

\begin{figure}[t]
\begin{lstlisting}[xleftmargin=0.7\columnsep]
/* Header */
package org.zoolu.sip.header;

/** SIP Status-line, i.e. the first 
 * line of a response message */
public class StatusLine {
	protected int code;
	protected String reason;
	
	/** Construct StatusLine */
	public StatusLine(int c, String r) {
		code = c;
		reason = r;
	}

	/** Create a new copy of the request-line */
	public Object clone() {
		return new StatusLine(getCode(), getReason());
	}

	/** Indicates whether some other Object
	 * is "equal to" this StatusLine */
	public boolean equals(Object obj){ 
		try {
			StatusLine r = (StatusLine) obj;
			if (r.getCode() == getCode()&&
			r.getReason().equals(getReason()))
				return true;
			else
				return false;
		} catch (Exception e) {
			return false;
		}
	}

	public int getCode() {
		return code;
	}

	public String getReason() {
		return reason;
	}

	public String toString() {
		return "SIP/2.0 " + code + " " + reason + "\r\n";
	}
}
\end{lstlisting}
\caption{
Original source code. A snippet from \texttt{bigbluebutton}'s \lstinline+StatusLine.java+. We use this
as a running example.}
\label{fig:ex} 
\vspace{-1em}
\end{figure}
\begin{figure}[t]
\createlinenumber{5}{6}
\createlinenumber{6}{7}
\createlinenumber{7}{8}
\createlinenumber{8}{9}
\createlinenumber{9}{10}
\createlinenumber{10}{11}
\createlinenumber{11}{15}
\createlinenumber{12}{16}
\createlinenumber{13}{17}
\createlinenumber{14}{20}
\createlinenumber{15}{21}
\createlinenumber{16}{23}
\createlinenumber{17}{24}
\createlinenumber{18}{25}
\createlinenumber{19}{26}
\createlinenumber{20}{27}
\createlinenumber{21}{28}
\createlinenumber{22}{29}
\createlinenumber{23}{30}
\createlinenumber{24}{31}
\createlinenumber{25}{34}
\createlinenumber{26}{35}
\createlinenumber{27}{36}
\createlinenumber{28}{39}
\createlinenumber{29}{40}
\createlinenumber{30}{43}
\createlinenumber{31}{44}
\createlinenumber{32}{47}
\begin{lstlisting}[xleftmargin=0.7\columnsep]
/* Header...*/
package org.zoolu.sip.header;

/** SIP Status-line, i.e. the first...*/
public class StatusLine {
	protected int code;
	protected String reason;	

	/** Construct StatusLine...*/
	public StatusLine(int c, String r) {...}

	/** Create a new copy of the request-line ..*/
	public Object clone() {...}

	/** Indicates whether some other Object...*/
	public boolean equals(Object obj){ 
		try {
			StatusLine r = (StatusLine) obj;
			if (r.getCode() == (getCode()&&
			r.getReason().equals(getReason()))
				return true;
			else
				return false;
		} catch (Exception e) {...}
	}

	public int getCode() {...}

	public String getReason() {...}

	public String toString() {...}
}
\end{lstlisting}
\caption{A summary of the file in \autoref{fig:ex} (left) which results from folding lines 1, 4--5, 11--14, 21--22, 31--33, 36-38 and 40-42. The ellipses indicate folded segments of code.}
\label{fig:ex2}
\vspace{-1em}
\end{figure}

\begin{itemize}
\item We introduce a novel autofolding method for source code summarization, called
TASSAL\footnote{https://github.com/mast-group/tassal}, based
on optimizing the similarity between the summary and the source file. Because of 
certain constraints among the folding decisions,
we formulate this method as a contiguous rooted subtree problem (\autoref{greedy}).
This is, to our knowledge, the first content-based autofolding method for code summarization.
\item To determine which non-essential regions should be folded, we 
introduce a novel topic model for code (\autoref{nlp_model}),
building on machine learning methods used in NLP \cite{haghighi2009exploring}, which separates tokens
according whether they best characterize their file, their project,
or the corpus as a while. This allows TASSAL summaries to focus on file-specific tokens.
\item We perform a comprehensive evaluation of our method on a set of popular open source projects
from GitHub (\autoref{experimental_setup}), and find that TASSAL performs better than simpler baselines
(\autoref{results}) at matching human judgements, with a relative error reduction of 28\%.
Furthermore, in a user study with experienced
developers, TASSAL is strongly preferred to the baselines.
\item We created a live demo of TASSAL \cite{fowkes2016tassal} to showcase how it can be used to summarize open-source Java projects on GitHub. Our demo can be found at \url{http: //groups.inf.ed.ac.uk/cup/tassal/demo.html} and a video highlighting the main features of TASSAL can be found at \url{https://youtu.be/_yu7JZgiBA4}.
\end{itemize}
More broadly, we hope that this work will aid program comprehension by turning code folding, perhaps an
overlooked feature, into a useful, usable and valuable tool. 

\section{Related Work}\label{related_work}
The application of NLP methods to the analysis of source code text
is only just beginning to be explored. Recent work has applied
language modelling \cite{allamanis2013mining,hindle2012naturalness,jacob2010code,lawrie2006what,nguyen13statistical},
natural language generation \cite{sridhara10summary,sridhara11automatically},
machine translation \cite{nguyen13lexical}, and topic modelling \cite{movshovitz2012natural} 
to the text of source code from large software projects. 
A main challenge in this area is to adapt existing NLP techniques to source code text. In contrast to natural languages,
programming languages are unambiguous, employ little redundancy, are meant to be interpreted literally, and consist of strictly structured text.
To exploit these features of the problem, we perform the summarization at the code block level, leveraging the fact that source code is syntactically unambiguous.

There is some existing work on the use of \textbf{code folding} (also known as code elision) to aid comprehension.
In particular, Cockburn \etal \cite{cockburn2003hidden} find that illegible elision of all method bodies in a class improves programmer efficiency in editing and browsing tasks.
Rugaber \etal \cite{rugaber2008conceptual} consider a conceptual model for manual folding, extending it to non-contiguous regions of code.
Kullbach \etal \cite{kullbach2001folding} develop the \texttt{GUPRO} IDE to aid in the comprehension of C preprocessor code via rule-based folding of macro expansions and file includes.
Also, Hendrix \etal develop the \texttt{GRASP} program comprehension tool, combining control structure diagramming with manual folding \cite{hendrix1998visual}.
Bragdon \etal \cite{bragdon2010code} perform code autofolding of long methods based on code block depth in their proposed Code Bubbles IDE.
However, they do not evaluate the effectiveness of the autofolding method on its own,
but rather as part of a larger UI. By contrast, we are the first to
quantitatively study and evaluate the autofolding problem directly.

The task of \textbf{natural language summarization} has been studied extensively \cite{sparck2007automatic}, mostly focusing on
\emph{extractive} summarization --- the problem of extracting the most relevant text segments from documents.
Source code identifiers (e.g., variable names) are information-rich and have been shown to be important for tasks such as feature 
location \cite{abebe2011effect,dit2013feature}. NLP techniques have been used on these identifiers for information retrieval tasks
such as automatically selecting labels for software artifacts \cite{de2013labeling}. Extractive summarization has also been applied
for the automatic summarization of bug reports \cite{mani2012ausum,rastkar2010summarizing}.

We are aware of only a few previous methods that consider the problem
of \textbf{code summarization}.
One of the first approaches is program slicing \cite{silva2012vocabulary,khoo2008path} which hides irrelevant LOC for a chosen program path -- essentially a very specific form of query-based summarization.
Program slicing focuses on the display of a path for a specific statement or variable of interest, and is not obviously applicable to the first look problem that we consider.
Most similar to our work are Haiduc \etal\cite{haiduc2010supporting,haiduc2010automated} and the follow up work by Eddy \etal\cite{eddy2013evaluating} and Rodeghero \etal \cite{rodeghero2014improving}, who also consider the problem of summarizing source code, particularly methods and classes, but in their work code fragments are summarized by a short list of keywords. For example, the \lstinline+equals+ method in \autoref{fig:ex} might be summarized by the list of terms (\lstinline+equals+, \lstinline+code+, \lstinline+reason+, \lstinline+Status+).  
McBurney \etal \cite{mcburney2014improving} take this idea further and present the keywords
in a navigable tree structure, with more general topics near the top of the tree.  
In our work, we summarize code \emph{with} code,
which we would argue has the potential to provide 
a much richer and more informative summary.

Also, Ying \etal \cite{ying2013code} consider the problem of summarizing a list of
code fragments, such as those returned by a code search engine.
They use a supervised learning approach at the level of lines of code.
Because they consider the results of code search,
their classifier uses query-level features, 
\eg, whether a line of code uses any identifiers that were present in the query.
This is a source of information that is not available in our problem setting.
In contrast, we target use cases in which the developer is skimming the source code
to get an overview of its operation,
rather than performing a directed keyword search.
Kim \etal\cite{kim2013enriching} develop a system that augments API documentation with code example summaries
but these are mined from the web and are therefore limited to APIs which have examples
already written for them --- our approach is applicable to \emph{any} source file.

On a more technical level, our folding-based summaries are distinguished
from this previous work in that our summaries are coherent with
respect to the programming language's syntax.
Indeed, Eddy \etal\cite{eddy2013evaluating} observe that developers prefer
summaries with a natural structure. Folding on code blocks
also enables us to retain method headers in the summary ---
identified by Haiduc \etal\cite{haiduc2010automated}
as highly relevant to developers and accounting for the
high scores of their best performing method.
Additionally, our
method leverages a multiple-project corpus during the summarization process,
which we exploit to identify tokens which are less characteristic
of a particular file.

In addition to extractive summarization methods, \textbf{abstractive summarization} techniques have also been used in software engineering research. 
Work in this area includes synthesis of API usage examples \cite{buse2012synthesizing}, extraction of API 
usage patterns \cite{wang2013mining,xie2006mapo}, and generation of natural language summaries for source code \cite{moreno2013automatic,sridhara10summary}.

The use of \textbf{topic models for source code} has also been studied in depth \cite{de2013labeling,gethers2011codetopics,savage2010topic,thomas2011mining,baldi2008theory,allamanis2013and,panichella2013effectively}. Marcus \etal\cite{marcus03lsi,marcus04concept} used Latent Semantic Indexing (LSI)\cite{deerwester1990indexing} for identifying traceability links and concept location in source code. More closely related to our work, Haiduc \etal\cite{haiduc2010automated} used LSI as a content model for their keyword-based source code summarizer. In their follow up paper, Eddy \etal additionally used a hierarchical pachinko allocation model (hPAM)\cite{mimno2007mixtures},
a family of generative topic models that build on Latent Dirichlet Allocation (LDA) \cite{blei2003latent} with a hierarchical topic structure. McBurney \etal \cite{mcburney2014improving} used the hierarchical document topic model (HDTM) \cite{weninger2012document} for their structured keyword-based summaries.
Note that HDTM is not the same as the topic model we propose: our model discerns file-specific tokens,
leveraging the hierarchical structure present in the code,  
whereas HDTM infers a tree that represents similarities between methods.  
Also, Movshovitz \etal \cite{movshovitz2012natural}
successfully used LDA and link-LDA\cite{erosheva2004mixed} for predicting class Javadoc comments from source file text.

We cast autofolding 
as an instance of the general problem of \textbf{selecting an optimal subtree} given a certain budget.
This problem has been studied theoretically by Guha
\etal\cite{guha1999efficient}, who propose a dynamic programming solution, but
this is only pseudo-polynomial time, and so is unlikely to scale well in
practice.

\section{Problem Formulation}\label{problem_formulation}
Our aim in this paper is to summarize source code so that
it conveys the most important aspects of its intended function. We envisage our proposed 
Tree-based Autofolding Software Summarization ALgorithm (TASSAL) being embedded
in a programming language IDE and providing real-time summaries to the user of
selected files.
The summarization could be useful at multiple levels, ranging from a single
source file to an entire corpus. For the
purposes of this paper, we will focus on the Java programming language as it is a
popular, high-level, platform-independent language. However, since TASSAL works entirely
with the source code's Abstract Syntax Tree (AST), it can be applied to {any} programming language for which an AST can be defined.

The target use case we envisage for TASSAL is that of a developer not familiar with
a project wishing to obtain an overview of a given file. For example, a developer who is considering using a new project on
GitHub might like to get an overview of the algorithms used in each file of the project.
We call this the \emph{first-look problem}.
The first-look problem is in contrast to tasks
such as debugging and code reviewing for which a more focused
summary may be desirable; we leave these other tasks to future work. 

The outline of TASSAL is as follows: TASSAL takes as input a
set of source files along with a desired compression ratio (\ie level of summarization) and outputs a
summary of each file where uninformative regions of code have been
folded (see \autoref{fig:ex2} for an example). In order to achieve
this TASSAL first selects the AST locations to obtain suitable regions to fold
(\autoref{code_summarization}). It then applies a source code language model to
each foldable region. The aim of this model is to identify, for every source file,
which tokens specifically characterize the file, as opposed to project-specific or Java-generic tokens that are not as informative for understanding the file.
To this end, we develop a scoped topic model for source code
(\autoref{nlp_model}),
which we apply to rank how informative each code region is
to its enclosing file.
Using this ranking TASSAL then leverages an optimization
algorithm to determine the most uninformative regions to fold while achieving
the desired level of compression. This is a novel optimization procedure that takes
the structure of the code into account (\autoref{greedy}).

\subsection{Problem Definition}\label{code_summarization}

Most modern IDEs already have extensive support for folding \emph{specific} code regions as well as the ability to fold regions based on user-specfied \emph{rules}. IDEs with support for automatically folding regions based on their \emph{location} have also been proposed \cite{bragdon2010code}. But to the best of our knowledge the problem of automatically determining which regions to fold based on their \emph{content} is novel. When we say that we \emph{fold} a source code region we mean that the region is replaced by a one
line summary and a symbol indicating that the region was folded.
We define the \emph{autofolding problem} as that of choosing a set of code regions to fold, 
such that the total length of the folded file as a fraction of the original is
below a user-specified compression ratio, and the remaining, unfolded, text captures the most important aspects of the file in question. Autofolding can be seen as a special case of \emph{extractive summarization}.

To encourage intuitive summaries, we let the
system perform folding only on \emph{code blocks}
(regions of source code delimited by \lstinline+{+ , \lstinline+}+), \emph{comment blocks} (regions 
delimited by \lstinline+/*(*)+ , \texttt{\color{mygreen}*/}), and \lstinline+import+ statements. 
We call these the \emph{foldable regions} of the code.
Our reasoning for this is that it is a summary many programmers are familiar
with as these are the regions
that can be manually folded in the majority of modern IDEs and text editors.
Moreover, code blocks are natural units for extractive summarization
since they take advantage of the code structure specified by the programmer.
However, since our approach works within the code's AST, it can be trivially
extended to fold \emph{any} contiguous region of interest. For example, in our implementation we have added optional features 
to allow autofolding of \emph{line comments}, \emph{fields} and
\emph{method signatures}. Also, it would be a trivial extension to allow statements, or a carefully designed subset thereof, to be folded.
In keeping with the manual folding conventions in IDEs, the one
line summary we display for a folded region consists of the first 
non-empty line of the code block, then an ellipsis, and finally the right delimiter of the region (see \autoref{fig:ex2}).

\begin{figure}[tb]
\centering
\includegraphics[width=0.47\textwidth]{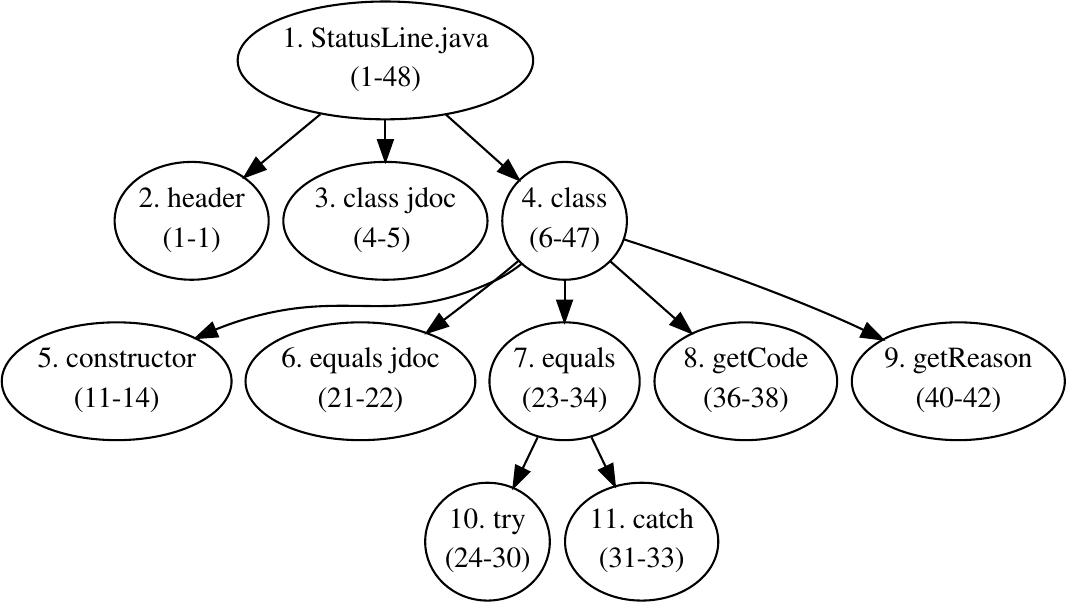}
\caption{Partial foldable tree constructed for \lstinline+StatusLine.java+ (\autoref{fig:ex}).
Numbered breadth-first with labels denoting block types and line numbers in brackets,
 \cf the source code snippet in \autoref{fig:ex}. Note that we have omitted some nodes for clarity.}
\label{fig:tree}
\vspace{-10pt}
\end{figure}

We formalize the autofolding problem by using the AST representation of the source code \cite{eclipseJdt}.
Given a program's AST,
we define the program's \emph{foldable nodes} as those AST nodes which
correspond to a foldable region of code. By starting at the root of an AST and
sequentially extracting all foldable nodes, we construct a directed
\emph{foldable tree}, containing just
the AST nodes we are interested in. \autoref{fig:tree} shows a partial foldable
tree for the running example.

Constructing a foldable tree enables us to formulate the summarization problem mathematically as finding the best
contiguous rooted subtree that takes up no more than a predefined number of lines of source code (LOC).
That is to say, we unfold all nodes in the best subtree and fold the remaining nodes in the tree.
Note that we require the tree to be rooted and contiguous as otherwise this would lead to confusing
situations where we would have a deep node present in the summary with no context.
We will describe the precise formulation in the next two sections.

\subsection{Content Model}\label{nlp_model}
In order to determine which nodes of the foldable tree should be unfolded, we
require
a content selection method for choosing the best nodes to retain in the summary. 
Intuitively, one would like to retain the most informative
nodes and a natural approach, as in text summarization, is to tokenize the node
text and select the nodes with the most representative tokens.
For this reason we make use of a \emph{topic model},
an extension of the TopicSum model \cite{haghighi2009exploring}.
The high-level idea is to extend a standard topic model, latent Dirichlet allocation, with topics that are specific
to particular projects, files and methods, thus allowing the model to identify which tokens are predominantly
used in a specific file or project.
Other types of probabilistic models that have been used for code are not well suited for this task.
For example, \emph{$n$-gram
language models} can learn only local patterns of nearby tokens, whereas we require a global model of the distribution of code across files.
An n-gram model has no way to identify whether the usages of a token are concentrated within a single file.

\boldpara{Tokenization}
A first idea would be to have one token in the topic model for each Java token
in the code. However, some tokens (\eg
operators and delimiters) are not informative about the program content and
identifier names have substructure that we wish to exploit.
For example, in \autoref{fig:ex}, the \lstinline+getCode+ method name
is closely related to the \lstinline+code+ member
variable, which becomes apparent to the topic model only if it
is split into two tokens \lstinline+get+ and \lstinline+code+.
For these reasons, we preprocess the Java tokens before incorporating
them into the topic model.
Given a code block, we first tokenize it into a set of Java tokens
using standard tools for the Java programming language.
Then we remove all tokens except for identifiers,
\ie, programmer assigned names of variables, methods, classes, \etc
Finally, we convert each of the identifiers into a new set of tokens
by splitting on camel case and underscores,
and converting to lowercase
(\eg, \lstinline+FooBarBaz+ becomes three tokens
\lstinline+foo+, \lstinline+bar+, and \lstinline+baz+,
as would \lstinline+foo_bar_baz+).
Additionally, we include the text of all comment blocks
in the topic model, splitting the comment text based on words,
again applying the identifier splitting procedure
on any comment tokens that contain camel case or underscores.
Let the \emph{vocabulary} $\mathcal{V} = \{t_1, \ldots t_T \}$ be the
set of all unique tokens that occur in the corpus. We use the term \emph{lexical item} $t \in \mathcal{V}$ to refer to elements in the vocabulary (which
can occur multiple times in the corpus as different tokens).

We do not use a stoplist because we expect that
the set of appropriate stop words for program text
would be different to those for natural language text.
Instead, our topic model identifies
background words automatically.

\boldpara{Vector Space Model (VSM)}
The VSM is a standard method in information retrieval \cite{manning:irbook}, 
in which documents are represented by continuous-valued vectors,
and similarity is measured by metrics such as the cosine similarity
between such vectors.  To apply this idea to summarization, we compare a vector
representing a source file with a vector representing a summary,
and find the closest match.  More specifically, for a file $f$, let 
$\bv_f$ be the log term frequency (tf) vector; this is vector containing
the log frequency of each token in the file. Similarly, for a potential summary
$\bu$ we can define a log-tf vector $\bv_\bu$. Then if we have a set of potential
summaries of the file $f$, we can choose the summary that maximizes
\begin{equation}
\sigma(\mathbf{u}) = \text{csim}(\bv_f, \bv_\bu), \label{eq:tfcost}
\end{equation}
where $\text{csim}$ denotes the cosine similarity.
One potential disadvantage of this method is that $\bv_f$
includes many generic tokens that are used throughout the project or multiple projects.
Therefore we next turn to methods that specifically identify such generic tokens.

\boldpara{Topic Model}
Now we describe an approach to identifying file specific tokens
based on a topic model. Because topic models are less familiar to a software engineering
audience, we will explain the model in some detail; although the mathematics of this section
may seem complex to some readers, in fact, the model that we employ 
is a straightforward extension to the basic latent Dirichlet allocation (LDA) model \cite{blei2003latent,steyvers2007probabilistic}.
A topic model \cite{blei12topic} is a type of statistical model over documents,
that represents each document as a combination of \emph{topics},
which are groups of words that tend to occur together.
Formally, each topic is modelled as a probability distribution $\phi_k$ over lexical items and can be viewed as a vector of length $T$ where each entry $\phi_{kt}$ is a probability. A document is modelled as a probability distribution
$\theta_d$ over $K$ topics; this can again be represented as a vector
where each element $\theta_{dk}$ is a probability that
represents how important topic $\phi_k$ is to document $d$.
Given a document $d$ with tokens $\wB_d = (w^{(d)}_1 \ldots w^{(d)}_{N_d})$,
we seek to infer topic assignments $z^{(d)}_i$ for each token $w^{(d)}_i.$
Each topic assignment $z^{(d)}_i \in \{1 \ldots K\}$ is an indicator
variable that specifies which topic was responsible for
generating the token $w^{(d)}_i$.  We use $\zB_d = (z^{(d)}_1 \ldots z^{(d)}_{N_d})$
to denote a vector of topic assignments to all the tokens in document $d$.

Topic modelling typically follows the paradigm of Bayesian statistics.
One first defines a probability
distribution $P(\phi_1 \ldots \phi_K,
\theta_1 \ldots \theta_D, \zB_1 \ldots \zB_D, \wB_1 \ldots \wB_D)$ that describes how the topics and documents would be distributed if all modelling
assumptions were correct.  Often the easiest
way to describe such a model is to consider
an algorithm that samples from it. In the case of a topic model,
this will be an algorithm that samples documents.
%
Then, when we receive a corpus 
of documents $\wB_1 \ldots \wB_D$, we infer topics for that corpus
by computing the posterior distribution $P(\phi_1 \ldots \phi_K,
\theta_1 \ldots \theta_D, \zB_1 \ldots \zB_D | \wB_1 \ldots \wB_D).$ The posterior distribution is the conditional distribution over
the quantities we don't know, given the ones that we do;
it is uniquely determined by the joint probabilistic model
and the laws of probability.  Given samples from the posterior distribution, we  estimate
each topic $\phi_k$ by averaging over the samples.

TopicSum \cite{haghighi2009exploring} is a \emph{scoped} topic model
that extends LDA to handle topics at multiple levels.
In TopicSum, each topic can be one of three kinds: a) a probability distribution over words that is 
local to a single document, b) a distribution over words that is local to a subcollection of related 
documents (such as all articles from the New York Times), c) a distribution over background words that 
is available to all documents in the corpus. The intention is that the background topic models stop 
words, the document subcollection topic represents significant
content and the document topic very specific document words.

We adapt TopicSum to source code by defining
a set of scopes that are appropriate for program text.
Specifically, in our model we consider five different kinds of topic: one topic $\phi_f$ for each file
$f$, one topic $\phi_p$ for each software project $p$ 
and three background topics shared across projects. Although the
model isn't aware of this, in practice we find that the three background topics
correspond to common Java tokens ($\phi_{B_J}$), common Javadoc comment
tokens ($\phi_{B_D}$) and common header comment tokens ($\phi_{B_H}$).
This model fits well into our summarization procedure
because it separates out which tokens are characteristic
of general Java code, of a specific project and of a specific file,
so that when we generate the summary we can focus
on preserving the file-specific tokens rather than the
generic Java tokens.

\begin{figure}[tb]
\centering
\includegraphics[width=0.34\textwidth]{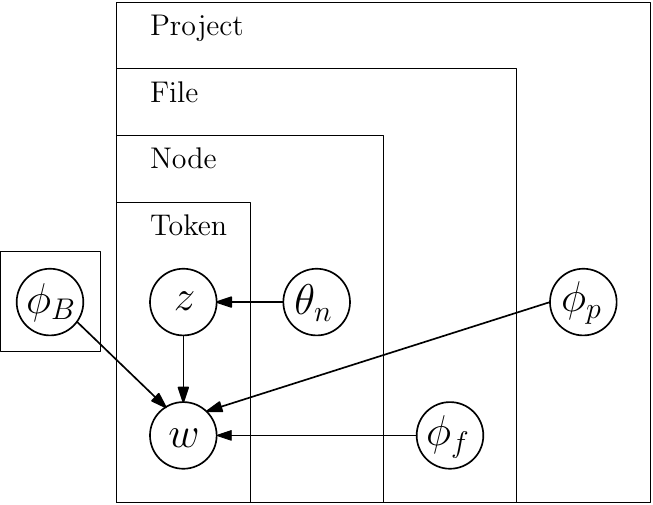}
\caption{Graphical model depiction of the TASSAL content model using plate notation
(we refer readers unfamiliar with such notation to the tutorial \cite{buntine1994operations}).
The plates denote repeated groups of variables. 
}
\label{fig:topicModel}
\end{figure}

We describe the model by providing
a procedure to sample from it.
In our model, a file $f$ in project $p$ is
generated as follows: each token $w_{i}^{(nfp)}$ in every
foldable node $n$ of the file is chosen from a specific topic $\phi_{z_{i}^{(nfp)}}$, where
the topic assignment $z_{i}^{(nfp)}$ is selected according to the distribution over
topics $\theta_n$ in the node $n$. That is, 
we have the following generative procedure: 
\begin{enumerate}
 \item Choose token distributions $\phi_k \sim \text{Dirichlet}(T,\beta_k)$ for
topics $k \in \{B_J,B_D,B_H,p,f\}$.
\item Choose topic distribution $\theta_n \sim
\text{Dirichlet}(K,\alpha\mathbf{m})$ for node $n$.
\item For each token $w_{i}^{(nfp)}$ in node $n$:
\begin{enumerate}
\item Choose a topic $z_{i}^{(nfp)} \sim \text{Categorical}(K,\theta_n)$.
\item Choose a token $w_{i}^{(nfp)} \sim \text{Categorical}(T,\phi_{z_{i}^{(nfp)}})$.
\end{enumerate}
\end{enumerate}
Here $\alpha\mathbf{m}$ and $\beta_k$ denote hyperparameters for the prior
distributions of topics and tokens, \ie the initial topic and token
assignments. Readers familiar with topic models will
recognize this as a simple scoped extension of LDA
\cite{steyvers2007probabilistic}.
\autoref{fig:topicModel} is a graphical illustration of our model using plate
notation, we refer readers unfamiliar with
such notation to the tutorial \cite{buntine1994operations}.

To estimate the topics $\phi_k$,
we need to compute the posterior distribution over the topic assignments $z$.
Unfortunately, as in most topic models, this distribution is intractable, so we
use a popular approximation called \emph{collapsed Gibbs sampling}.
Collapsed Gibbs sampling  is a stochastic iterative procedure, which does not make explicit reference to the parameters $\theta_d$ and $\phi_k$, 
which at each iteration returns samples of
topic assignments $z_{i}^{(nfp)}$ for each token $w_{i}^{(nfp)}$ in the corpus.
It can be shown that such successive samples approximate the
posterior \emph{marginal} distribution over topic assignments
\cite{steyvers2007probabilistic}, marginalizing out $\theta_d$ and $\phi_k$. 
We omit the details of the algorithm for space, but it is a simple
 extension of the one used for LDA \cite{steyvers2007probabilistic}.
For the purposes of this paper, the sampler can be thought of simply as a black box that
outputs a
topic assignment $z_{i}^{(nfp)}$ for each token in the corpus.

Once we have samples of $z_{i}^{(nfp)}$, it is easy to compute an estimate of each topic distribution $\phi_k$,
which is what we will need for summarization.
These can be estimated using the maximum a posteriori (MAP) estimate  \cite{steyvers2007probabilistic}
\begin{equation}\label{eq:map}
 \hat{\phi}_{kt} = \frac{N_{t|k}+\beta_k}{\sum_{t=1}^T \left( N_{t|k}+\beta_k
\right)},
\end{equation}
where $N_{t|k}$ denotes the number of times the lexical item $t$ has been assigned to topic $k$
in the entire set of samples $\{ z_{i}^{(nfp)} | \forall i, n, f, p \}$.
In other words, we simply compute the proportion of times that topic $k$ is used to generate each lexical item $t \in \mathcal{V}$.

Additionally, unlike the original TopicSum model which used fixed
hyperparameters $\alpha \mathbf{m},\boldsymbol\beta$,
we incorporate efficient hyperparameter optimization, namely
MacKay and Peto's method \cite{mackay1995hierarchical} with histogram based
computation of the necessary counts (see \cite{wallach2008structured}
for details).

\subsection{Optimization Method}\label{greedy}

Given a content model, we now need an algorithm
that extracts the most relevant summary by selecting suitable AST nodes to unfold in the foldable tree.
We propose an iterative greedy optimization algorithm
to extract the most relevant rooted contiguous subtree
from the foldable tree given constraints on the subtree size.
At each iteration of the greedy procedure, we have a candidate summary, \ie, a
subtree we have gathered so far and decided to include in the summary. We then
consider each additional node of the foldable tree to decide whether it
is relevant and should also be included.


We need to compare a candidate summary to the original file $f$.
For a VSM, cosine similarity is a standard measure.
For topic models, a suitable information-theoretic measure for this
similarity is the Kullback-Leibler (KL) divergence, which measures the
difference between two probability distributions. To
convert the candidate summary into a probability distribution,
we divide the number of times each token $w$ occurs in the summary over the total number of tokens
in the summary, \ie by using the \emph{empirical unigram distribution}.
Using our topic model, we can estimate the file token
distribution $\hat{\phi}_f$ via \eqref{eq:map}.
We can therefore assign the node under consideration a score
based on the KL divergence between the corresponding file token
distribution $\hat{\phi}_f$ and the empirical unigram distribution of the candidate
summary. Intuitively, we want our summary to contain tokens that characterize the file rather than tokens
that are common elsewhere in the corpus.


Formally, for each node $i$ let $u_i \in \{0,1\}$ indicate whether it is unfolded, with $1$ corresponding to true (\ie
unfolded) and let $\mathbf{u}$ be the vector containing all the $u_i$ (so that $\mathbf{u}$ contains $u_i$ at position $i$). We define a score $\sigma(\bu)$ for a candidate summary $\bu$. For the VSM, we use $\sigma(\bu)$ as \eqref{eq:tfcost}
whereas for the topic model we use the score
\begin{equation}\label{eq:score}
\sigma(\mathbf{u}) = -KL(\hat{\phi}_f|P_{\mathbf{u}})
\end{equation}
where $\hat{\phi}_f$ denotes the file token
distribution, $P_{\mathbf{u}}$ the empirical unigram distribution over tokens in the candidate summary
and $KL(\cdot|\cdot)$ the KL divergence. Foldable regions
with no tokens are assigned a score of $-\infty$ for consistency, which
means that they are never unfolded.

We can now formulate the autofolding summarization problem as finding the
optimal rooted contiguous subtree. 
Requiring the  subtree to be rooted and contiguous
ensures that all nodes in the summary are presented within their syntactic context. 
Suppose we wish to summarize a file using a line-level compression ratio 
of $p\%$, \ie, we would like to compress the file to $p\%$ of its original size in LOC.
We can then define $L_{max}$, the maximum number of lines of code that are allowed in the summary, as
$L_{max} \deq (p/100)L_0$
where $L_0$ denotes the number of LOC in the original file. 
Note that since we are folding on a block level, a line-level compression ratio of $p\%$ does not mean 
that $p\%$ of the blocks are compressed, in fact for small files all the blocks
are often folded at $50\%$ compression.

Moreover, folding on a block level also means that in practice the specified line-level compression ratio will never be achieved exactly, but instead the returned summary will always be slightly shorter.
In certain situations, it may be desirable to allow the specified compression ratio to be slightly 
exceeded, such as when a slightly larger summary would have many more relevant terms.
In our approach, we do not handle this tradeoff in the optimization method and the target compression 
ratio is taken as a hard maximum.  It would be interesting to modify the user interface so as to indicate
when a slight increase in compression ratio would lead to large increase in the estimated quality of the summary, 
for example, by visualizing the score $\sigma(\mathbf{u})$ alongside a slider that controls the compression ratio,
but we leave this to future work.

Let $G = (V,E)$ be the foldable tree, that is, a directed tree with a set of AST nodes
$V=\{1,\dotsc,N\}$ consisting of the aforementioned foldable regions and $E$ a
set of \emph{directed} edges between the nodes, where it is understood that $(i,j) \in
E$ means that $i$ is the parent node of $j$ (\cf \autoref{fig:tree}). Furthermore, for a node $i \in V$
let $L_i$ denote the number of LOC underneath node $i$. We then define $C_i$, 
the cost of unfolding node $i$ as the number of LOC \emph{unique} to node $i$, \ie,
underneath node $i$ but not any of its children. Formally,  
\begin{equation}\label{eq:cost}
 C_i \deq (L_i-1) -\sum_{j : (i,j) \in E} (L_j-1).
\end{equation}
The first line of a node is
never folded (\cf \autoref{fig:ex2}), hence the minus one.
Let $\sigma(\mathbf{z})$ denote the score obtained from the summary
nodes $u_1,\dotsc,u_N$ as defined in \eqref{eq:tfcost} or \eqref{eq:score}. The optimal rooted contiguous subtree problem is
\begin{subequations}
\label{eq:prob}
\begin{align}
\max_{\mathbf{u}} &\; \sigma(\mathbf{u}) \label{eq:prob1}\\
\text{s.t. } & \sum_{i=1}^N C_i u_i \le L_{max} \label{eq:prob2} \\
& u_i \ge u_j \qquad \text{ if } (i,j) \in E \label{eq:prob3}\\
& u_i \in \{0,1\} \qquad \forall \, i \in V. \label{eq:prob4}
\end{align}
\end{subequations}
That is, we unfold the nodes which maximise the total score \eqref{eq:prob1} subject
to staying below the maximum allowed LOC \eqref{eq:prob2} and retaining a rooted
contiguous subtree \eqref{eq:prob3}.

As the score \eqref{eq:score} is nonlinear we approximate this problem using a greedy approximation algorithm:
that is we unfold the next available node that will give the maximum score \eqref{eq:score} per
cost \eqref{eq:cost} increase, honoring the cost constraint \eqref{eq:prob2} and unfolding any folded parent nodes \eqref{eq:prob3}. That is, starting from all the nodes being folded (\ie $\mathbf{u}=\mathbf{0}$),
iteratively choose the node that maximizes
\[
 \sigma(\mathbf{u}+\mathbf{e_i})/C_i
\]
while $\sum_{i=1}^N C_i u_i \le L_{max}$. Here $\mathbf{e_i}$ denotes the $i$-th unit vector, which is $1$ at position $i$ and $0$ everywhere else.

\section{Experimental Setup}\label{experimental_setup}

\begin{table*}[t]
\small
\centering \begin{tabular}{llrrrrr} \toprule
 Project  & Description & LOC & Methods & Classes & Forks & Watchers \\ \midrule
 \textsf{storm}  & Distributed Computation System & 59,827 & 5,740 & 761 & 1,416
& 7,471 \\
 \textsf{elasticsearch} &  REST search engine & 518,905 & 32,077 & 4,990 & 1,283
& 5,246 \\
 \textsf{spring-framework} & Application Framework & 798,249 & 47,214 & 8,395 &
1,774 & 2,568 \\
 \textsf{libgdx} & Game Development Framework & 334,706 & 33,821 & 2,651 & 1,844
& 2,243 \\
 \textsf{bigbluebutton} &  Web Conferencing System & 105,315 & 6,364 & 852 &
1,602 & 969 \\
 \textsf{netty} &  Network Application Framework & 160,579 & 10,324 & 1,267 &
927 & 2,304 \\
\bottomrule \end{tabular}
\caption{The top Java projects on GitHub, used in the current work. Ordered by
popularity.}
\label{tbl:projtable}
\end{table*}

In this section, we describe how we obtain a gold standard summary on real code.
The gold standard allows automatic evaluation,
which is the de facto standard in NLP summarization \cite{lin2004rouge}.
Automatic evaluation is a standard technique in modern
artificial intelligence research. It is vital because it leads to a rapid development cycle in which we can compare many more differing
algorithms than would be possible with user studies alone.
In particular, it enables us to perform a comprehensive evaluation of TASSAL in \autoref{results}.

\boldpara{Data}
To evaluate the performance of TASSAL we obtained the source code for the
top six Java projects on the popular GitHub service for hosting
open source code repositories.
The top six projects were determined by a popularity score, 
which is the sum of the number of forks and the number of watchers, where each is separately normalized to have zero mean and unit variance. We selected the projects with the highest score that were greater
than 100,000 KB in size as of December 1st 2013. These are given in 
\autoref{tbl:projtable} along with a brief description of their domains.  

\begin{figure}[t]
\centering
\includegraphics[width=0.43\textwidth]{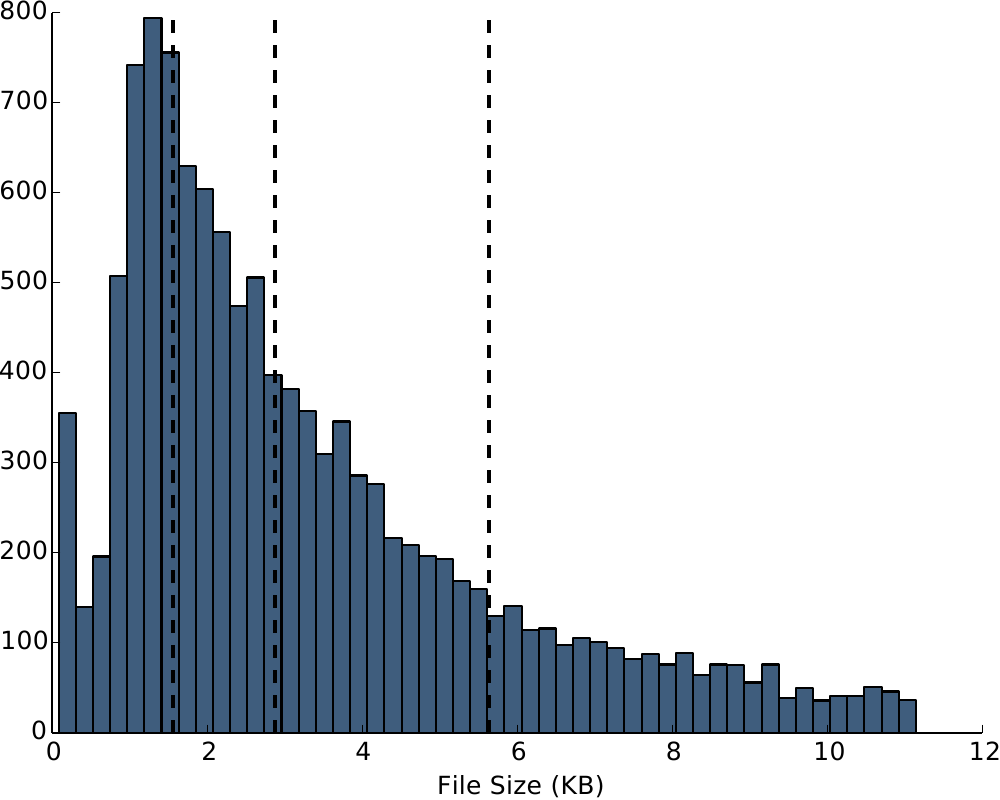}
\caption{Histogram of file sizes across the top projects (dashed vertical lines
denote quartiles).}
\label{fig:hist}
\end{figure}

For each of these six projects, we divided the project files into quartiles by file size, 
and sampled four files from each quartile, to obtain a total of 96 files ($12,347$ LOC),
each one of which
was annotated by two human annotators. Only \texttt{.java} files were 
considered (excluding the \texttt{package-info.java} files, which only contain package-level documentation).
\autoref{fig:hist} shows the distribution of file sizes across the top
projects.

\boldpara{Annotation Procedure}
Human annotators were given guidelines prior to performing the
annotation. Annotators were asked, for each source
code file, to manually fold the file in Eclipse until they reached
a compression ratio of 50\% and save their work.
The annotators were asked to fold regions that would be least useful to a developer
who was reading the file for the first time in order to understand its overall purpose,
reflecting our interest in the first-look problem.
A compression ratio of 50\% was chosen as, on average, it gave the best balance between providing a reasonable summary whilst not eliding all the interesting details of the underlying source code.
Although 50\% may not seem like a dramatic compression, in fact many of the
remaining lines are block headers or blank (\cf the running example in
\autoref{fig:ex2}). We found that on average across all 96 files in the annotated dataset, half of the LOC remaining unfolded in the file were blank or block headers.
Moreover, when we ran TASSAL on the dataset, we found that
$22\%$ of files had their top-level class entirely folded due to the fact that unfolding \emph{any} nodes in the top level class would have resulted in exceeding the $50\%$ compression ratio.

Annotators were allowed to browse the full source code of the
project while annotating each file. We used two experienced Java
developers as annotators, who each independently annotated 
the entire data set.
We performed our annotation
prior to the development of
our summarization system so it was impossible for the annotators to
unconsciously favour the system's output in their judgements.

Annotators were asked to fold regions until they thought the file conveyed 
the most important aspects of its function (or equivalently provided a good 
overview of its purpose to a programmer unfamiliar with it) 
with the following constraints: Annotators were asked to always fold import statements and header
comments (such as copyright notices). Empty and one-line blocks were also
folded by default. Setters and getters along with other I/O methods were
asked to be folded, unless they contained core logic of the code. Similarly, 
annotators were asked to fold commonly overridden and overloaded methods (\eg \texttt{toString()} methods) 
unless they provide sufficient new information about the functionality of
the code. Finally, Javadoc and block comments were left unfolded if they were informative and
succinctly explained the function of the associated class or method. Comments
whose text spanned only one line were folded by default.

As for potential threats to validity, the first concern that could be raised is whether
it is indeed possible to expect different annotators who are generating a reference
summary to have similar ideas about what constitutes a good summary. To address this concern,
we report in the next section the inter-annotator agreement at the line level,
which reflects substantial agreement among the annotators.
Second, one could raise the concern that different types of developers,
or developers who are considering different tasks, would require different
types of summaries. We would agree with this concern:
it is certainly true that developers who are familiar with the projects could
favor a different type of summary than the annotators. However, recall that in this study, our target use
case focuses on the first-look problem, \ie developers who are new to the
projects, for which we would argue our annotators are good representatives.
Finally, note that annotators were asked to fold regions whereas TASSAL unfolds regions, 
however this is merely for mathematical convenience and the two formulations are entirely equivalent.

\begin{figure}[t]
\centering
\includegraphics[width=0.45\textwidth]{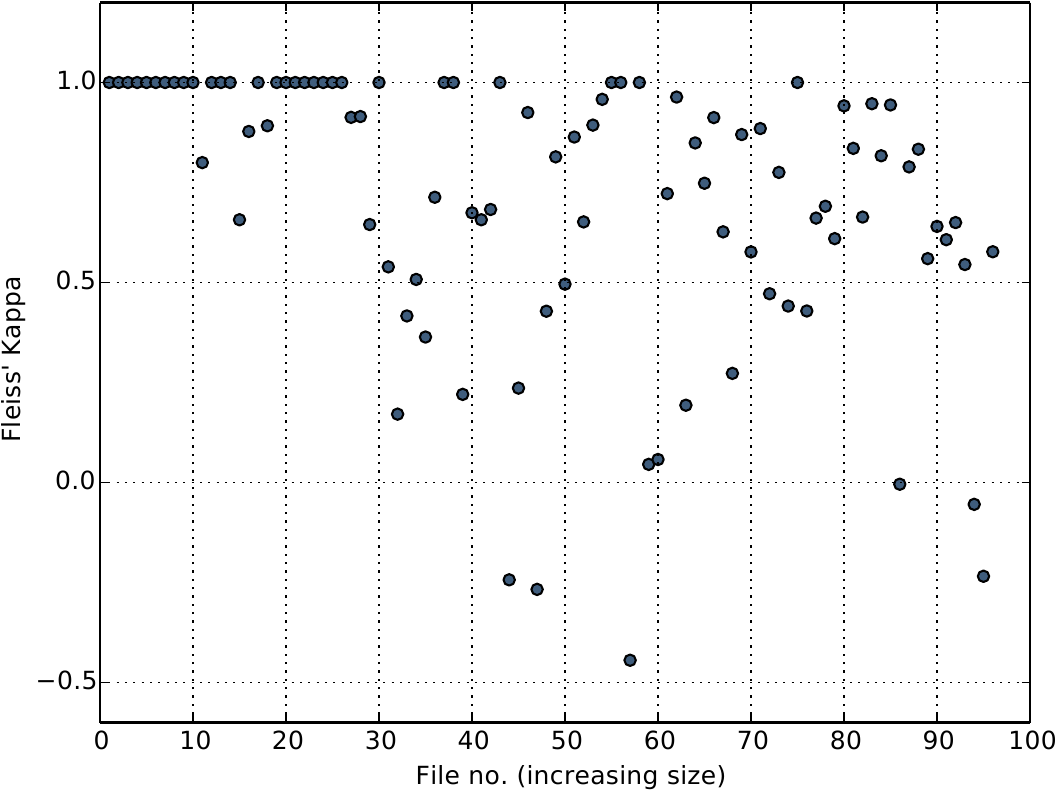}
\caption{Line-level inter-annotator agreement for the 96 source files at a compression ratio of $50\%$.}
\label{fig:agreement50}
\end{figure}

\boldpara{Annotation Statistics}
As autofolding is a new task, we need to verify that the task
is well-defined.
We therefore calculated the line-level agreement between annotators
and found that it was substantial,
with a Fleiss' Kappa value of 0.71 at 50\% compression,
averaged across all files. \autoref{fig:agreement50} shows the 
line-level inter-annotator agreement for each of the 96 files at a compression ratio of
50\% and one can see that for the majority of files we obtain substantial
agreement between the two annotators.
We also found that the line-level annotation agreement for non-trivial nodes (i.e. nodes that are not \emph{block comments} or \emph{import statements}) was moderate with a Fleiss' Kappa of 0.60.

\section{Results}\label{results}
In this section, we evaluate the performance of TASSAL
against our annotated test set from \autoref{experimental_setup}. 
We begin by training and assessing the quality of TASSAL's underlying topic
model
before moving on to a comprehensive evaluation of TASSAL itself.
As our method is fully unsupervised, we use all 96 annotated files from 
the six projects in our corpus for evaluation.

\boldpara{Topic Model}
We train TASSAL's topic model on all 12,093 \texttt{.java} files from the six
projects in our corpus.
This enables the topic model to automatically discern common coding patterns,
such as common libraries, and thus allows TASSAL to recognize unimportant code.
We run the topic sampling algorithm for 5,000 iterations,
performing hyperparameter optimization every 10 iterations, to infer our trained model.

\begin{table*}[tb]
\small \centering
\texttt{%
\begin{tabular}{lllllll} \toprule
 \multicolumn{3}{c}{\textrm{Background}} & \multicolumn{2}{c}{\textrm{Project}}
& \multicolumn{2}{c}{\textrm{File}} \\ \cmidrule(r){1-3} \cmidrule{4-5}
\cmidrule(l){6-7}
 \textrm{Java} & \textrm{header comments} & \textrm{Javadoc comments} &
spring-framework & bigbluebutton & DataSourceUtils & QuaLsp \\ \cmidrule(r){1-3}
\cmidrule{4-5} \cmidrule(l){6-7}
 get & the & the & bean &  sip & connection & lsp \\
 string & license & a & org & org & con & j \\
 value & or & to & test & log & holder & constants \\
 name & under & of & context & it & source & k \\
 type & you & link & springframework & event & data & ld8 \\
 object & distributed  & for & exception & gnu & synchronization & mode \\
 i & of  & is & request & listener & isolation & tmp \\
 set & 0 & this & factory & message & order & wegt \\
 exception & is & and & get & public & level & index \\
 map & 2 & code & class &  general & close & m \\
\bottomrule
\end{tabular}
}
\caption{The top ten tokens in each topic-type as found by our topic model.}
\label{tab:tokens}
\end{table*}

To demonstrate that the topic model learns to distinguish between project-specific and file-specific tokens, we show the top ten tokens in each of the five topic
types (Java background, Javadoc background comment, header background comment, project and file) for two projects and two files in \autoref{tab:tokens}.
Note that background tokens which are very common in a file/project can also appear among the top ten
file/project tokens.
One can see that these are representative of their respective topics:
the Java background topic contains many commonly used terms in Java such as \texttt{get},
\texttt{set}, \texttt{map} and \texttt{object}.
The background comment topics on the other hand contain
common English stop words such as \texttt{the}, \texttt{a}, \texttt{or}
and one can clearly see the distinction between the topics, \ie,
words that commonly appear in headers (such as \texttt{license}, \texttt{distributed})
are found in the header background comment topic whereas
those that commonly appear in Javadoc comments (\texttt{link}, \texttt{code}) are found in the Javadoc background comment topic.
Looking at the topics for the \texttt{spring-framework} project
one can see that it contains the fully resolved project name \texttt{org.springframework} as well
as common project-specific tokens such as \texttt{bean}, \texttt{context} and \texttt{factory}.
Similarly, the \texttt{bigbluebutton} project contains the specific tokens \texttt{sip},
\texttt{event}, \texttt{message}
and \texttt{gnu}. Looking at selected files from both projects, the \texttt{DataSourceUtils} file topic contains tokens specific to the
function of that file, \eg \texttt{connection}, \texttt{data},
\texttt{synchronization} and \texttt{isolation}.
Finally, the \texttt{QuaLsp} file topic, a codec implementation, contains the very function-specific tokens \texttt{lsp} and \texttt{ld8}.

This illustrates the quality of our proposed topic model
and suggests that it can distinguish file- and project-specific tokens from those that are common across all Java projects.
It also raises the exciting possibility that our topic model is robust enough for wider applications and not merely restricted to summarization.

\boldpara{Baselines}
To provide a comprehensive performance comparison of TASSAL, we
also evaluate three alternative baseline systems that represent more 
na\"{i}ve approaches for summarizing source code. All baselines 
start from a fully folded tree and gradually unfold nodes, making local
decisions, until they reach the required compression ratio.
If a node is to be unfolded, all of its parent nodes are also unfolded as well.
The baselines are:
\begin{description}
\item \textbf{Shallowest} unfold the shallowest available node first,
choosing randomly if there is more than one. This would unfold node 1 in
\autoref{fig:tree} first, followed by either node 2, 3 or 4, \etc
\item \textbf{Largest} unfold the largest available node first,
as measured by the number of tokens, breaking ties randomly.
In \autoref{fig:tree}, this would unfold the class block (node 4)
first, followed by the class Javadoc (node 3), \etc
\item \textbf{Javadoc} first unfold all Javadoc comments (in random
order) and then fallback at random to an available node, unfolding methods last.
This would unfold the Javadoc nodes 3, 6 in \autoref{fig:tree} first
and the method nodes 5, 8, 9 last (7 would already be unfolded as
it is the parent of 10 and 11). 
\end{description}
Each of the baselines represents a possible assumption that we can make
about summarizing source code. The Largest baseline assumes that the largest nodes are more valuable in a summary, the Shallowest baseline is representative of the folding approach used in the Code Bubbles IDE \cite{bragdon2010code} and the Javadoc baseline is representative of the current defaults of IDEs such as Eclipse. While we are aware that these baselines are rather simple methods, this reflects the fact that the autofolding problem has not received much attention in the research literature --- to our knowledge, there simply do not exist more advanced methods in the literature for us to compare against.

To further verify our annotation procedure, we implemented the basic guidelines followed by the annotators as an additional system: 
\begin{description}
\item \textbf{Guidelines} unfold getters/setters, 
I/O methods (\texttt{read|write|load|save.*}), block comments, import statements and empty blocks last in that order, unfolding other node types first in random order. In \autoref{fig:tree}, this would unfold nodes 1, 3--7, 10, 11 first in random order followed by the getter nodes 8, 9 and the header node 2.
\end{description}
This allows us to get some good insight into how much additional content-sensitive folding was done by the annotators after these basic annotation guidelines were followed: we found that only $27\%$ of all nodes were forced to be folded by the guidelines, meaning that $73\%$ of all nodes that were folded in the gold standard require human judgement.    

\boldpara{Performance}
To assess the performance of TASSAL and the
baselines, we used the annotated test set provided by each of the two annotators
as our gold standard in turn, averaging the results across both annotators.
In this way we can measure whether the output of the summarizer matches human
judgements.

For each file in the test set, we treated the folding problem as a binary
classification, classifying each line of code in a foldable region as either
unfolded (positive) or folded (negative), excluding one-line foldable regions (which are indistinguishable since all foldable regions are folded to one line).
This enabled us to calculate the average accuracy, precision and recall of our
summarizer across all test files and also $F_1$ as the harmonic mean of the
average precision and recall. 

We compared both TASSAL with the topic model and TASSAL with VSM (which we denote as TASSAL VSM). 
The resulting average performance metrics at a compression
ratio of 50\% are given in \autoref{tab:lines50}.
As one can clearly see from the results, TASSAL (with topic model)
outperformed all the systems, by a margin of about 10\% when compared against
the best performing baseline (Javadoc). TASSAL VSM on the other hand performed poorly, losing to the Javadoc baseline. We also performed the same analysis on
each foldable region of code, \ie classifying each foldable region as either
unfolded or folded. The resulting average performance metrics are given in
\autoref{tab:folds50} and one can see that TASSAL is once again the best
performing system.

\begin{table}[tb]
\small\centering
\begin{tabular}{lcccccc} 
\toprule
 & Accuracy & F1 & Precision & Recall\\ \midrule
\textbf{TASSAL} & \textbf{0.77} & \textbf{0.75} & \textbf{0.74} & \textbf{0.76}\\
TASSAL VSM* & 0.65 & 0.61 & 0.61 & 0.60\\
Javadocs*  & 0.68 & 0.65 & 0.64 & 0.66\\
Shallowest*  & 0.65 & 0.62 & 0.60 & 0.63\\
Largest*  & 0.60 & 0.56 & 0.56 & 0.57\\
\specialrule{0.1pt}{0.5pt}{1.3pt}
Guidelines  & 0.74 & 0.71 & 0.71 & 0.72\\
\bottomrule
\end{tabular}
\caption{\emph{Per-line} evaluation statistics for the summarizers evaluated on
\emph{all nodes} at a compression ratio of 50\%. Averaged across annotators
and all test files, ordered best first. *significantly different from TASSAL ($p < 0.05$).}
\label{tab:lines50}
\end{table}

\begin{table}[tb]
\small\centering
\begin{tabular}{lcccccc} 
\toprule
 & Accuracy & F1 & Precision & Recall\\ \midrule
\textbf{TASSAL} & \textbf{0.69} & \textbf{0.55} & \textbf{0.53} & \textbf{0.56}\\
TASSAL VSM* & 0.57 & 0.43 & 0.39 & 0.50\\
Javadocs & 0.60 & 0.46 & 0.42 & 0.52\\
Shallowest & 0.61 & 0.46 & 0.43 & 0.50\\
Largest & 0.65 & 0.47 & 0.47 & 0.46\\
\specialrule{0.1pt}{0.5pt}{1.3pt}
Guidelines  & 0.62 & 0.50 & 0.47 & 0.54\\
\bottomrule
\end{tabular}
\caption{\emph{Per-node} evaluation statistics for the summarizers evaluated on
\emph{all nodes} at a compression ratio of 50\%. Averaged across annotators and
all test files, ordered best first. *significantly different from TASSAL ($p < 0.05$).}
\vspace{-12pt}
\label{tab:folds50}
\end{table}

We are also interested in the performance of TASSAL on \emph{non-trivial}
foldable regions, \ie foldable regions that are not Java \emph{block comments}
or \emph{import statements}. 
Note that \emph{Javadoc
comments} are considered non-trivial as these often contain usage and
implementation details. 
We therefore performed the same analysis as above on non-trivial regions
at both a line- and node-level. The results are given in Tables
\ref{tab:lines_nontriv50},\ref{tab:folds_nontriv50} and one can see that TASSAL
remains the best performing system.
One can clearly see the advantage of using a topic model here, as TASSAL consistently outperforms TASSAL VSM which exhibits similar performance to the baselines. 

Turning our attention to the Guidelines system, we find that it performs very well as expected 
(we are after all following and evaluating on the annotation guidelines), outperforming all the
baselines and TASSAL VSM. This not only further validates our annotated dataset (demonstrating that our annotators did indeed follow the basic guidelines given to
them), but also demonstrates that TASSAL is indeed able to match human judgements: $28\%$ of 
the nodes folded by the Guidelines system at $50\%$ compression are required to be folded by 
the annotation guidelines, meaning that $22\%$ of the folded nodes require human 
judgement. As TASSAL clearly outperforms the Guidelines system in Tables \ref{tab:lines50}--\ref{tab:folds_nontriv50}, we can conclude that, despite being an unsupervised system, TASSAL is able, to some degree, to match the intuitive human
judgements of the annotators. Conversely, this also shows that TASSAL VSM is either unable to match the basic 
guidelines or the human judgements or both, clearly demonstrating the need for a more
sophisticated content model.     

As for statistical significance, we calculated two-tailed $p$-values using
\texttt{sigf} \cite{sigf2006} for the $F_1$ scores on both gold standards.
The difference between TASSAL and the baselines is significant
at the line-level for all nodes and non-trivial nodes ($p<0.05$). At
the node-level, the difference between TASSAL and the baselines is not significant. 

As a further test of whether the proposed summaries are plausible, we consider
several classes of methods that are likely to be
uninteresting, and verify that TASSAL usually folds them.
We show the percentage of times header comments, imports, constructors, getters,
setters and other generally uninteresting pattern-based method types
were folded at $50\%$ compression in \autoref{tab:methods}. 
TASSAL folds these methods in most cases. 
Furthermore, when such methods are included in the summary, these exceptional
methods turn out to be qualitatively more interesting,
as we verify by manual examination.
We show example snippets in Figures \ref{fig:snippets1}--\ref{fig:snippets3}. As one can see, the unfolded methods
tend to exhibit unusual or non-standard behaviour in the method body, so much so
that they cannot be easily summarized by their signature alone. This lends further evidence
to the credibility and usefulness of our summarization approach. 

\begin{table}[tb]
\small\centering
\begin{tabular}{lcccccc} 
\toprule
 & Accuracy & F1 & Precision & Recall\\ \midrule
\textbf{TASSAL} & \textbf{0.62} & \textbf{0.66} & \textbf{0.57} & \textbf{0.77}\\
TASSAL VSM* & 0.54 & 0.55 & 0.50 & 0.60\\
Javadocs*  & 0.54 & 0.57 & 0.50 & 0.66\\
Shallowest*  & 0.56 & 0.57 & 0.52 & 0.63\\
Largest*  & 0.54 & 0.53 & 0.50 & 0.57\\
\specialrule{0.1pt}{0.5pt}{1.3pt}
Guidelines  & 0.59 & 0.62 & 0.55 & 0.72\\
\bottomrule
\end{tabular}

\caption{\emph{Per-line} evaluation statistics for the summarizers evaluated on
\emph{nontrivial nodes} at a compression ratio of 50\%. Averaged across
annotators and all test files, ordered best first. *significantly different from TASSAL ($p < 0.05$).}
\label{tab:lines_nontriv50}
\end{table}

\begin{table}[tb]
\small\centering
\begin{tabular}{lcccccc} 
\toprule
 & Accuracy & F1 & Precision & Recall\\ \midrule
\textbf{TASSAL} & \textbf{0.71} & \textbf{0.59} & 0.57 & \textbf{0.61}\\
TASSAL VSM* & 0.61 & 0.51 & 0.49 & 0.53\\
Javadocs & 0.63 & 0.52 & 0.49 & 0.56\\
Shallowest & 0.65 & 0.51 & 0.51 & 0.51\\
Largest & 0.70 & 0.52 & \textbf{0.60} & 0.46\\
\specialrule{0.1pt}{0.5pt}{1.3pt}
Guidelines  & 0.65 & 0.56 & 0.54 & 0.59\\
\bottomrule
\end{tabular}
\caption{\emph{Per-node} evaluation statistics for the summarizers evaluated on
\emph{nontrivial nodes} at a compression ratio of 50\%. Averaged across
annotators and all test files, ordered best first. *significantly different from TASSAL ($p < 0.05$).}
\vspace{-12pt}
\label{tab:folds_nontriv50}
\end{table}

\begin{table*}[t]
\small\centering
\begin{tabular}{ccccccccccccccc} \toprule
 \textsf{header} & \textsf{import} & \textsf{construct} & \textsf{get} & \textsf{set} & \textsf{put} & \textsf{is} & \textsf{has} & \textsf{read} & \textsf{write} & \textsf{add} & \textsf{remove} & \textsf{contains} & \textsf{clear} & \textsf{reset} \\ \midrule
 80\% & 70\% & 83\% & 83\% & 88\% & 100\% & 91\% & 100\% & 75\% & 80\% & 100\% & 80\% & 100\% & 100\% & 83\% \\
\bottomrule \end{tabular} 

\caption{The percentage of times specific types of node in the test set (header comments, import statements, constructors and method names starting with the listed keywords) were
\emph{folded} at $50\%$ compression by TASSAL.}
\label{tab:methods}
\end{table*}

\begin{table}[t]
\small\centering
\begin{tabular}{lcccc} 
\toprule
 & \multicolumn{2}{c}{All Nodes} &
\multicolumn{2}{c}{Nontriv.\ Nodes} \\ 
\cmidrule{2-3} \cmidrule(l){4-5}
 & Per-line & Per-node & Per-line & Per-node \\
\cmidrule(r){1-1} \cmidrule{2-3} \cmidrule(l){4-5}
\textbf{TASSAL} & \textbf{0.86} & \textbf{0.79} & \textbf{0.57} & \textbf{0.79} \\
TASSAL VSM & 0.72 & 0.56 & 0.51 & 0.63 \\
Javadocs & 0.81 & 0.71 & 0.53 & 0.71 \\
Shallowest & 0.77 & 0.69 & 0.54 & 0.71 \\
Largest & 0.73 & 0.74 & 0.55 & 0.79 \\
\specialrule{0.1pt}{0.5pt}{1.3pt}
Guidelines & 0.82 & 0.66 & 0.54 & 0.69 \\
\bottomrule
\end{tabular} 

\caption{Area under the curve (AUC) for the receiver operating characteristic (ROC) curves of the summarization systems as the compression ratio is varied. }
\vspace{-12pt}
\label{tab:auc}
\end{table}

It is also evident from the results that the Shallowest and Largest baselines perform 
poorly and it is not difficult to see why. The shallowest nodes tend
to be top-level code blocks defining methods and classes, which rarely contain core
logic directly. Rather, the core logic tends to be nested in children of top-level blocks, such as \lstinline+if-else+
statements, \lstinline+for/while+ loops and \lstinline+try-catch+ blocks.
The Largest nodes
may contain a substantial amount of code but
rarely the core logic as the number of tokens is a bad indicator of code importance.
To see this, consider header comments which contain many tokens (words) 
often stating the code copyright, 
or common class methods such as the equals method in 
\autoref{fig:ex} which perform routine functions yet contain blocks with many identifiers.
The Javadoc baseline, on the other hand, represents the de facto summarization method 
currently used in IDEs and therefore performs much better as one would expect.

We compared the performance of the summarization systems at a range of
compression ratios (effectively treating it as a threshold) against the gold
standard at the fixed compression ratio of $50\%$. The standard approach to evaluating 
a binary classifier with a threshold is via the area under the curve (AUC) values 
for its receiver operating characteristic (ROC) curve. An
ROC curve plots the recall (equiv.\ true positive rate) 
against the fraction of false positives out of the negatives (false positive rate) 
at various thresholds (compression ratios). 
Essentially the
higher above the diagonal a binary classifier is, the better it performs and a
good measure for this is the AUC.
Intuitively, the AUC measures the quality of the ranked list produced by the system, i.e., where all the lines/nodes are ranked by how likely the system is to fold them, whereas the F scores we used earlier measure only the performance at a single compression ratio.
The AUC for each of
the summarizers is given in \autoref{tab:auc}. Once again, we can see that
TASSAL outperforms all the baselines with an AUC of $0.86$ when evaluated per-line on all nodes in the gold standard.

Finally, it should be noted that TASSAL has a very fast runtime, 
needing less than five seconds of CPU time to summarize an average
file in the dataset on a 2.66GHz Core 2 Quad machine (note that this excludes training the topic model which, in an implemented system, we assume is handled in an offline preprocessing step). 

\boldpara{Developer Study}
We conducted a developer study to test whether content-based autofolding methods produces better summaries than non-content based methods.
We asked developers to rate the summaries produced by our best
content-based method, TASSAL using the topic model, and the summaries
the three non-content baselines and one randomly chosen annotator at a compression
ratio of $50\%$. We recruited six experienced developers for our study,
separate
from the annotators who created the gold standard. All
were recently graduated computer science masters students with an average $5.3$ years Java
programming experience and $4$ years industry programming experience.

\begin{table}[t]
\small\centering
\begin{tabular}{lcccc} 
\toprule
 & \multicolumn{2}{c}{Conciseness} &
\multicolumn{2}{c}{Usefulness} \\ 
\cmidrule{2-3} \cmidrule(l){4-5}
Summary & Mean & St.\ dev.\ & Mean & St.\ dev.\\
\cmidrule(r){1-1} \cmidrule{2-3} \cmidrule(l){4-5}
Gold & 3.34 & 1.03 & 3.33 & 1.04 \\
\textbf{TASSAL} & \textbf{3.27} & \textbf{1.01} & \textbf{3.18} & \textbf{0.97}
\\
Javadocs* & 3.07 & 1.03 & 2.69 & 1.09 \\
Shallowest* & 2.97 & 1.05 & 2.50 & 1.15 \\
Largest* & 3.08 & 1.07 & 2.67 & 1.06 \\
\bottomrule
\end{tabular} 

\caption{Mean and standard deviation averaged across developer ratings for
summaries produced by the four summarization systems and a randomly chosen gold
standard at a compression ratio of $50\%$. *significantly different from Gold
and TASSAL ($p < 0.05$).}
\vspace{-12pt}
\label{tab:devstudy}
\end{table}

\begin{figure*}[t]
\begin{minipage}{.5\textwidth}
\begin{lstlisting}[xleftmargin=0.1\columnsep,numbers=none]
public SocksInitResponse(SocksAuthScheme authScheme) {
    super(SocksResponseType.INIT);
    if (authScheme == null)
        throw new NullPointerException("authScheme");
    this.authScheme = authScheme;
}

public ChannelGroupException(
		Collection<Map.Entry<Channel, Throwable>> causes) {
	if (causes == null)
		throw new NullPointerException("causes");  
	if (causes.isEmpty())
		throw new IllegalArgumentException("causes must be 
            non empty");
	failed = Collections.unmodifiableCollection(causes);
}
\end{lstlisting}
\caption{Snippets of constructors that were \emph{unfolded} by TASSAL.}
\label{fig:snippets1}
\;\;\,
\begin{lstlisting}[xleftmargin=0.1\columnsep,numbers=none]
public <T> T getOption(ChannelOption<T> option) {
	if (option == SO_TIMEOUT)
		return (T) Integer.valueOf(getSoTimeout());
	return super.getOption(option);
}

public void set(final ContactID c) {
	indexA = c.indexA;
	indexB = c.indexA;
	typeA = c.typeA;
	typeB = c.typeB;
}
\end{lstlisting}
\caption{Snippets of getters/setters that were \emph{unfolded} by TASSAL.}
\label{fig:snippets2}
\end{minipage}
\begin{minipage}{.5\textwidth}
\begin{lstlisting}[numbers=none]
public void readFrom(StreamInput in) throws IOException {
	super.readFrom(in);
	if (in.getVersion().onOrBefore(Version.V_0_90_3)) 
		in.readBoolean(); // refresh flag
	full = in.readBoolean();
	force = in.readBoolean();
}

public static void writeVLong(DataOutput stream, long i) 
  throws IOException {
    if (i >= -112 && i <= 127) {
        stream.writeByte((byte) i);
        return;
    }
    int len = -112;
    if (i < 0) {
        i ^= -1L; // take one's complement'
        len = -120;
    }
    long tmp = i;
    while (tmp != 0) {
        tmp = tmp >> 8;
        len--;
    }
    stream.writeByte((byte) len);
    len = (len < -120) ? -(len + 120) : -(len + 112);
    for (int idx = len; idx != 0; idx--) {
        int shiftbits = (idx - 1) * 8;
        long mask = 0xFFL << shiftbits;
        stream.writeByte((byte) ((i & mask) >> shiftbits));
    }
}
\end{lstlisting}
\caption{Snippets of readers/writers that were \emph{unfolded} by
TASSAL.}
\label{fig:snippets3}
\end{minipage}
\end{figure*}

To this end, we randomly selected four of the six projects and five of the
annotated files from each project for the study, resulting in 20 files in
total. For every file, developers were presented with each of the five
possible summaries in random order and asked to rate the conciseness and
usefulness of each summary on a five-point Likert scale (higher is better). Developers were allowed
to browse the full source code of each project during the study.

We show the average ratings across all six developers in
\autoref{tab:devstudy} along with the average standard deviations. One can
see that summaries produced by TASSAL score around $0.2$ points
higher on conciseness and $0.5$ points higher on usefulness
than the three baselines. Moreover, TASSAL is only $0.07$ points
lower on conciseness and $0.15$ points lower on usefulness
than the gold standard summary. We performed ANOVA on the developer conciseness
and usefulness ratings for the different summaries and found that
the difference between TASSAL and the baselines was significant ($p < 0.05$) as
denoted in \autoref{tab:devstudy} whereas the difference between TASSAL and the
gold standard was indeterminate.

In a follow up questionnaire we asked the developers to summarize the measures
they used to rate each summary. The developers generally favoured the same
criteria that we identified when creating the annotation guidelines for the
gold standard in \autoref{experimental_setup}. That is to say, the developers
preferred not to see accessor and housekeeping methods but felt it was
important to show method/class Javadoc comments (as reported in previous
code summarization studies \cite{haiduc2010automated,eddy2013evaluating}).

In short, the study results clearly show that our summarization system
is not only preferred by developers over the baselines, it is almost as good as
the summary produced by an expert annotator. 

Some of the developers also offered interesting qualitative feedback about 
their experience with TASSAL, even though we did not explicitly ask them for this information. 
Developer 2 felt that in order to understand a project better, he would like to see ``some 
kind of graph of class relationships'' from which he could determine ``which class is the most 
important, which package is the most important, how each of the classes fits in holistically 
and how often some methods or classes are called or created'', 
in order to see the big picture before diving deeper into individual classes. 
Developer 5 thought that it would be helpful to include the ability to unfold method comments   
and bodies ``so the person browsing [the summary] could dig in deeper in a way that suited 
them''. This ability for the user to selectively unfold folded regions of code was subsequently 
implemented in our live demo of TASSAL (see \autoref{demo}). Developer 5 also preferred to
 understand the intention of a piece of code first and if they ``had some doubts'' then 
 take a closer look, lending further support for the need for a tool such as TASSAL.

\section{Web Demo}\label{demo}

\begin{figure*}[t]
\centering
\includegraphics[width=\textwidth]{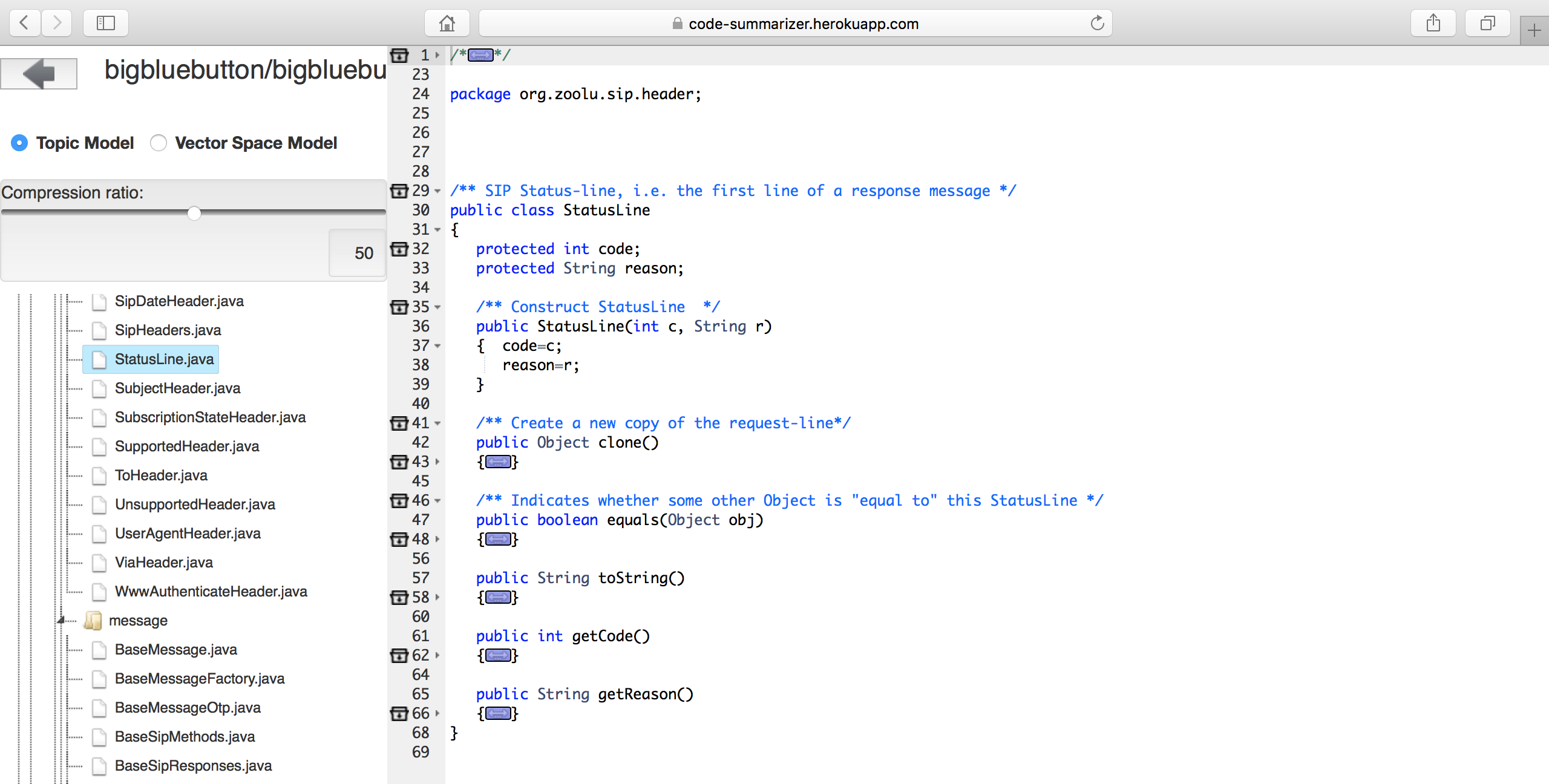}
\caption{A screenshot of our source code autofolding tool being used to summarize \texttt{StatusLine.java} from \texttt{bigbluebutton}.}
\label{fig:screenshot}
\end{figure*}

We created a live demo of TASSAL \cite{fowkes2016tassal} using the Play Framework (\url{https://www.playframework.com}) to showcase how it can be used to summarize open-source Java projects on GitHub (however note that TASSAL can summarize the source code of any Java project). Our demo of TASSAL can be found at \url{http://groups.inf.ed.ac.uk/cup/tassal/demo.html} and a video highlighting the main features of TASSAL can be found at \url{https://youtu.be/_yu7JZgiBA4}.
 
A screenshot of the demo is shown in \autoref{fig:screenshot} and as one can see from the figure, the basic layout of the demo is very simple. On the left hand side is a tree view showing all the Java source files for a user-selected project on GitHub. Upon clicking on a source file, the remainder of the screen uses the Javascript-based ACE code editor (\url{https://ace.c9.io}) to show a summary of the file where less informative code regions have been folded. The user can adjust the conciseness of the summary using the compression ratio slider at top-left, ranging from viewing the complete file (0\%) to folding all the foldable regions (100\%). As a sanity check, the fold icons ($\includegraphics[height=2ex]{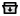}$) to the left of the line numbers denote the code regions that were marked as folded by TASSAL. Note that while TASSAL is able to fold fields, we did not find a satisfactory way to implement this in ACE and therefore omitted it from the demo (however the fold icons for fields are still displayed).

If the user wishes to unfold a folded region, they can do so by clicking on the symbol denoting the fold ($\includegraphics[height=2ex]{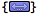}$) and conversely, they can fold any foldable region by clicking on the down arrow ($\blacktriangledown$) to the right of the line numbers as is standard in modern editors. One can see from the example (\texttt{StatusLine.java} from the \texttt{bigbluebutton} project displayed at 50\% compression) that the header has been folded, as have the \lstinline+toString+, \lstinline+getCode+, \lstinline+getReason+, \lstinline+clone+ and \lstinline+equals+ methods, \ie Java boilerplate code --- precisely the less salient code regions. Note also, how by \emph{folding} the less informative regions the code \emph{remains readable and navigable} and \emph{no information is lost}. This is not true of other summarization approaches to source code \cite{eddy2013evaluating,haiduc2010supporting,haiduc2010automated,khoo2008path,mcburney2014improving,rodeghero2014improving,silva2012vocabulary}.

As for the choice of language model, using TASSAL with the topic model will in general produce better summaries (\autoref{results}), however training a topic model is too expensive for an interactive system. Therefore, we train the topic model in advance on a small set of projects and cache the topic assignments. If the user requests summaries of GitHub projects for which we have not run the topic model, we fall back to the the VSM model. When both language models are available, the user can toggle between them by means of a radio button in the top-left corner.

\section{Discussion and Conclusions}\label{conclusions}
We presented a novel fully unsupervised approach for extractive source
code summarization, proposing that the folding procedure
common to IDEs can serve as the basis of an automatic summary.  We formulated this autofolding problem as an optimal subtree problem
on the source code's AST. 
Our method incorporates a novel topic model for source code that identifies which
tokens are most relevant to their enclosing files and projects.
Our evaluation demonstrates that our summarizer outperforms
several baselines, achieving an error reduction of $28\%$, is favoured by
experienced developers and even outperforms methods used as standard in modern
IDEs.
Furthermore, our live demo showcases how our summarizer can be used to summarize open-source Java projects on GitHub.

As with all automatic summarization systems, TASSAL is not perfect and indeed any algorithm that performs summarization based solely on file topics cannot be, since language, be it natural or programming, has a rich semantic structure. However, by looking in detail at the summaries produced, we can gain an insight into situations where TASSAL fails to perform optimally and what improvements, if any, we can make. 
A detailed examination of the file summaries used in the developer study brought to light the following cases:  
(a) \emph{Unfolding obvious comments}. Once a Javadoc comment containing very standard content (parameter and return types) and little additional information (stating that the method is a constructor) was unfolded. It is very difficult for for an automatic system to determine what constitutes an obvious comment as this requires deeper semantic information. 
(b) \emph{Unfolding getters/setters with obvious names}. Once a getter with an obvious name was unfolded because it had an interesting method body. Again, it is difficult for for an automatic system to determine what constitutes an obvious name as this also requires more semantic information. 
(c) \emph{Deciding whether a Javadoc or associated class/method body provides a better summary}. Once a class Javadoc provided a better summary of the class than the body of the class, which was itself interesting. Once again, an automatic system would require substantial semantic information to correctly decide this.

In future, we would like to extend our approach to generate targeted summaries for
specific software engineering tasks such as bug localization or code review
as well as investigating the possibility of folding at a statement (rather than block) level while maintaining a coherent summary.  
We would also like to explore the idea, suggested by one of the study participants, of creating a higher-level overview of class relationships that would allow a developer to see the bigger picture before delving into summaries of specific classes. 
Finally, in response to the shortcomings in TASSAL's current content model, we would like to see whether a more sophisticated content model based on deep learning would be better able to capture the semantic information in the source code. 
  
More broadly, NLP techniques for source code are only just beginning to be explored and have the potential for a much wider range of exciting applications from learning coding conventions to programming with natural language.

\section*{Acknowledgments}
This work was supported by the
Engineering and Physical Sciences Research Council (grant number EP/K024043/1)
and by Microsoft Research through its PhD Scholarship Programme.
We are also grateful to Rebecca Mason for allowing us to adapt her \texttt{TopicSum} implementation to source code and would like to thank Brian Doll for useful discussions.

\bibliographystyle{IEEEtran}
\bibliography{tse}

\end{document}